\documentclass[journal]{IEEEtran}

\usepackage{graphicx,cite,epsfig,amssymb,amsmath,subfigure,url,stfloats,latexsym}
\usepackage{bm}
\usepackage{algorithm}
\usepackage{algorithmic}
\usepackage{cases}
\usepackage{citesort}
\usepackage{color}
\usepackage{epstopdf}
\usepackage{amssymb}
\usepackage{cite}
\usepackage[caption=false,font=normalsize,labelfont=sf,textfont=sf]{subfig}
\usepackage{amsmath,lipsum}
\usepackage{cuted}

\newtheorem{theorem}{Theorem}

\newtheorem{lemma}{Lemma}

\newtheorem{corollary}{Corollary}

\begin{document}
	
	\title{Joint Security-Latency Design for Short Packet-Based Low-Altitude Communications}
	\author{
		Zeyin Wang,
		Di Zhang,~\IEEEmembership{Senior Member,~IEEE,}
		Shaobo Jia,
		Lulu Song,
		Yanqun Tang\vspace{-15pt}
		
		\thanks{This study was supported by the National Science Foundation of China under grant 62301502, U22A2001, the Henan Natural Science Foundation for Excellent Young Scholar under Grant 242300421169, 252300421224.}
		
		\thanks{Corresponding author: Di Zhang (E-mail:dr.di.zhang@ieee.org).}
		
		\thanks{Zeyin Wang, Lulu Song and Shaobo Jia are with the School of Electrical and Information Engineering, Zhengzhou University, Zhengzhou 450001, China (E-mail: eiezeyinwang@gs.zzu.edu.cn, ieshaobojia@zzu.edu.cn, lulu\_song@gs.zzu.edu.cn).}
			
		\thanks{Di Zhang is with the School of Intelligent Systems Engineering, Sun Yat-sen University, Shenzhen 518107, China, and also with the School of Electrical and Information Engineering, Zhengzhou University, Zhengzhou 450001, China (E-mail: dr.di.zhang@ieee.org).}
            
		\thanks{Yanqun Tang is with the School of Electronics and Communication Engineering, Sun Yat-sen University, Shenzhen 518107, China (E-mail: tangyq8@mail.sysu.edu.cn).}
	}

	\maketitle 
	\begin{abstract}
		In this article, a joint security and latency analysis of short packet-based low-altitude communications when the eavesdropper is close to the receiver is addressed. To reveal the impacts of the signal-to-noise ratio (SNR) and block-length on latency in communications, we propose a new metric named secure latency (SL) and derive the expressions for the effective secure probability (ESP) and the average SL. To minimize the average SL, different transmission designs are analyzed, in which the optimal solutions of SNR and block-length are provided. Numerical results validate our analysis and reveal the trade-off between reliability and security and the impacts of the block-length, SNR, and packet-generating rate on average SL, of which SNR and the block-length account for main factors. In addition, we find that the performance of SL can be enhanced by allocating less SNR.
	\end{abstract}
					\vspace{-4 pt}

	\IEEEpeerreviewmaketitle
			\vspace{-5 pt}
	\section{Introduction}	
	 Future wireless networks are anticipated to support ultra-reliable and low-latency communications (URLLC) in fifth generation (5G) or even the next generation URLLX (xURLLC) in six generation (6G), which enables many emerging low-altitude applications, e.g., unmanned aerial vehicle (UAV) delivery and intelligent transportation systems, where massive connected devices are distributed within a small area.  \cite{Xuewan_Chinacomm,Xuewan_IoTJ}. The shared characteristics of these low-altitude communications are latency-sensitive, information-less, but mission-critical, especially for the control and command type data\cite{zhuyao}. Given this, the transmissions with URLLC and xURLLC are carried out via short block-length codes, commonly called short-packet communications (SPC) \cite{Starris}. Departing from the traditional assumption of infinite block-length, transmission errors cannot be ignored in SPC, even if the transmission rate is less than the channel capacity. Therefore, the impact of short packet-length codes on reliability and latency should be carefully considered. To solve this problem, the bounds on the maximal achievable transmission rate for short block-length are derived \cite{Polyanskiy_Rate}. Based on this, extensive works are done, such as green communications \cite{Green}. 
	
	 On the other hand, retransmission is widely used to ensure reliability for SPC in current networks. However, the inherent broadcast nature of the wireless medium renders information security a vital concern in SPC system design \cite{zhuyao}. To address this issue, security design in the physical layer can be a feasible solution, e.g., covert communication \cite{Bash}, jamming \cite{Che}, etc. However, the interaction between reliability, latency, and security could have been neglected in previous studies, since all of them are key performance indicators. Especially in low-altitude communications, devices are close and their channels are correlated, resulting in a trade-off between reliability, latency, and security \cite{zhuyao}. Therefore, the traditional latency based on reliable transmission can no longer effectively describe the relationship between these aspects. Besides, to the best of our knowledge, there is scarcely any metric or research on reliability, latency, and security analysis jointly.
	
	 Motivated by the above discussions, in this article, we first introduce a model that can jointly characterize both security and latency for short packet-based low-altitude communications, where the eavesdropper is close to the receiver. Afterwards, we propose a new metric named secure latency (SL), derive the effective secure probability (ESP) and the average SL, and formulate the optimization problem to reveal the impacts of signal-to-noise ratio (SNR) and block-length on average SL. The optimal SNR and block-length designs to minimize the average SL are provided. In addition, numerical results demonstrate the trade-off between reliability and security and reveal that the security-latency performance can be enhanced by allocating less SNR.
	
	The rest of this article is organized as follows. The system model and preliminary are given in Section II. Section III presents the analysis and derivations. The optimal SNR and block-length are introduced in Section IV. The simulation results are shown in Section V, and the article is finally concluded by Section VI.
				\vspace{-7 pt}
	\section{System Model and Preliminary}
				\vspace{-5 pt}
	Considering that the legitimate users, say, Alice and Bob, are trying to transmit $D$ bits information that does not want to be detected by the eavesdropper, say, Eve. We assume each of them is equipped with a single antenna. Note that in short packet-based low-altitude communications, although Eve is close to Bob, they still are discernible. Since we focus on jointly investigating the security and latency of SPC, rather than beamforming, without loss of generality, we assume that the wireless channels from Alice to Bob and Eve are subject to additive white Gaussian noise (AWGN), and that the communication time is divided into slots $T$. When Alice transmits information of $L$ block-length to Bob in a time slot, the signal received by Bob can be written as 
	\begin{equation}
		y_b(i)=h_bx(i)+n_b,
	\end{equation}
where $i$ equals the index of the block-length, $x$ stands for the signal, $h_b$ and $n_b \sim \mathcal{N} (0,\sigma_b^2)$ represent the channel gain and noise between Alice and Bob, and the SNR at Bob is denoted as $\gamma_b$. Eve performs a statistical hypothesis test based on two circumstances during a time slot to determine whether Alice has transmitted a packet or not. The observed signals of Eve are given by\vspace{-5pt}
	\begin{equation}\resizebox{0.55\hsize}{!}{$
		y_e(i)=
		\begin{cases}
			n_e        \quad   \qquad \qquad \mbox{for} \quad H_0,  \\
			h_ex(i)+n_e \quad          \mbox{for} \quad H_1, 
		\end{cases}$}		\end{equation}
	where $H_0$ is the null hypothesis, $H_1$ is the alternative hypothesis, $h_e$ and $n_e \sim \mathcal{N}(0,\sigma_e^2)$ indicate the channel gain and noise between Alice and Eve, and the SNR at Eve is denoted as $\gamma_e$.
				\vspace{-20 pt}
	\subsection{Preliminary }\label{AA}
	For a determined information rate of $R=D/L$, decoding error probability is given as \cite{Polyanskiy_Rate}	
	\begin{equation}
		P_{e} \approx Q(\frac{C(\gamma)-R}{\sqrt{V(\gamma)/L}}),
	\end{equation}
	where $Q(\cdot)=\int_{x}^{\infty}\frac{1}{\sqrt{2\pi}}\exp(-t^2/2) dt$ equals a Q-function, $\gamma$ indicates the SNR, $C(\gamma)=\log (1+ \gamma)$ denotes the channel capacity, and $V(\gamma)= \gamma (2+\gamma)/(1+\gamma)^2$ defines the channel dispersion.
	
	When Eve tries to detect the transmission, it is inevitable to encounter two types of detection errors, i.e., missed detection and false alarm. Excluding the missed detection and the false alarm situations, according to Pinsker’s inequality \cite{Bash}, an upper bound of the detection probability is given as $\mathcal{V}_T(H_0,H_1) \leq \sqrt{\frac{D(H_0,H_1)}{2}}$, where $\mathcal{V}_T(H_0, H_1)$ is the total probability of correct detection for $H_0$ and $H_1$, and $D(H_0,H_1)=0.5L(\ln(1+\gamma)-\frac{\gamma}{\gamma+1})$ represents the relative entropy between $H_0$ and $H_1$. Following the inequality, we can formulate the detecting probability at Eve by the upper bound as \cite{Bash}
	\begin{equation}\resizebox{0.90\hsize}{!}{$
		P_{d}=\sqrt{\frac{D(H_0,H_1)}{2}}=\sqrt{\frac{L}{4}(\ln(1+\gamma_e)-\frac{\gamma_e}{\gamma_e+1})} .$}
	\end{equation}	
	
	\vspace{-10 pt}
	\section{ESP and SL Analysis}

	To guarantee a reliable transmission, retransmission mechanism is adopted in this article. Since Eve is close to Bob, SNR at Eve is also close to Bob's \cite{correlation}. So Bob can emulate Eve's detection through a likelihood ratio test based on two hypotheses and Eve's decoding situation, and then provide timely feedback signals.	
				\vspace{-10 pt}
	\subsection{ESP Definition}\label{AC}	
	In secure communication systems, one aims to achieve reliable transmission between Alice and Bob while preventing Eve from detecting the transmission or decoding the packet to ensure security, thereby maintaining a reliable and secure transmission. The probability of such a reliable and secure transmission is defined as ESP. All situations of the transmission are shown in Table~I.	
	\begin{table}[ht]
		\caption{All Cases of Effective Security }
		\centering	
		\begin{tabular*}{\hsize}{@{\extracolsep{\fill}}c c c c c}
			\hline
			&$(1-P_B)$   & $ P_{d} $      & $(1-P_E)$   &               \\
			case &decoding at Bob & detection at Eve & decoding at Eve &   Result     \\
			\hline
			1&$\times$        & $\times$       & $-$         & $\times$     \\
			2&$\times$        & $\bigcirc$     & $\times$    & $\times$     \\
			3&$\times$        & $\bigcirc$     & $\bigcirc$  & $\times$     \\
			4&$\bigcirc$      & $\times$       & $-$         & $\bigcirc$   \\
			5&$\bigcirc$      & $\bigcirc$     & $\times$    & $\bigcirc$   \\
			6&$\bigcirc$      & $\bigcirc$     & $\bigcirc$  & $\times$     \\
			\hline
		\end{tabular*}
		\label{table10}
	\end{table}
	 If Eve does not detect the transmission, he believes there is no packet and will not proceed with decoding operations, represented as "-". And "$\bigcirc$" is success while "$\times$" represents failure.
	As shown from the table, only for cases $4$ and $5$, the communication is reliable and secure, and the ESP in every transmission can be computed as
	\begin{equation}
		\begin{aligned}
		\mathit{P_{ESP}}&~= (1-P_B)(1-P_d)+(1-P_B)P_dP_E \\
		&~\triangleq(1-P_{B})(1-P_{d}(1-P_{E})),
	\end{aligned}
	\end{equation}	
	where $(1-P_{B})=1-Q(\frac{C(\gamma_b)-R}{\sqrt{V(\gamma_b)/L}})$ defines the probability of the packet reliably decoded by Bob, $P_{E}=Q(\frac{C(\gamma_e)-R}{\sqrt{V(\gamma_e)/L}})$ denotes the probability of the packet wrongly decoded by Eve, and $(1-P_{d}(1-P_{E}))$ represents the probability that Eve either wrongly detects the transmission or fails to decode the packet, indicating the security limitation. In addition, the effective secure rate with symbol rate $B$ can be written as $R_{ES}=BR(1-P_{B})(1-P_{d}(1-P_{E}))$.

	Since retransmissions are triggered when transmission fails, another factor influencing SL is the transmitting time. Because of the short block-length, the coding delay is much smaller compared to the transmission duration time. Therefore, each transmitting time equals the duration time, which can be denoted as $T$.	
\vspace{-8 pt}
	\subsection{SL Analysis}\label{CA}

	A reliable and secure transmission from Alice to Bob fails when decoding does not succeed on Bob or the transmission is successfully detected and decoded by Eve. If the transmission fails, Alice will retransmit the information in the following slots until success, and after that, Alice will wait for the next packet to transmit. To jointly characterize reliability, security, and latency for SPC, we introduce a new metric named SL, which means the time elapsed from when the packet is transmitted by Alice until it is decoded by Bob securely and reliably.
	
	\begin{figure}
		\centering\vspace{-5pt}
		\setlength{\abovecaptionskip}{0mm}
		\includegraphics[height=2cm,width= 4.8cm]{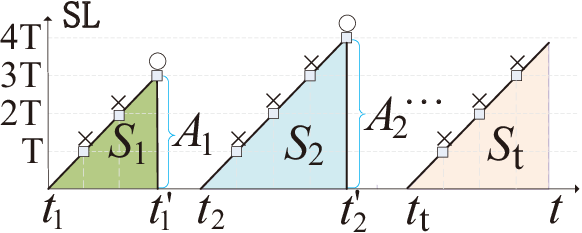}
		\label{fig:side:a}	
		\caption{Example of SL.}\vspace{-15 pt}	
	\end{figure}	 
	An example is shown in Fig. 1, where $t_i$ refers to the time when the $i$-th ($i=1 \dots n$) short packet starts to transmit, and $t_i'$ denotes the time when the packet is reliably and securely transmitted to Bob, and $T_{Wi}=t_i - t_{i-1}'$ represents the time for the transmitter to wait for the $i$-th packet after transmitting the $(i-1)$th packet, and there is a probability $\lambda$ generating a short packet in every time slot. $A_i=t_i'- t_i$ represents the SL of the $i$-th data packet, which means the elapsed transmitting time of the $i$-th reliable and secure packet. If the transmission succeeds, SL will immediately return to 0 and wait to transmit the next packet. Therefore, in every transmission, SL will be  
	\begin{equation}\resizebox{0.75\hsize}{!}{$
		SL_{i,k}=
		\begin{cases}
			0          \quad \ \qquad\qquad   \  \mbox{for} \quad \text{success},  \\
			SL_{i,k-1}+T  \quad           \mbox{for} \quad \text{failure},
		\end{cases}$}	
	\end{equation}
	where $SL_{i,k}$ means the latency of $k$-th transmission of the $i$-th packet. The average SL is obtained by averaging the SL over all time slots, representing the average time that information is securely and reliably received by Bob. For an interval (0, $\tau$), the average SL can be computed as 
	\begin{equation}\resizebox{0.85\hsize}{!}{$
			\overline{SL}=\lim\limits_{\tau\rightarrow\infty}\frac{1}{\tau}\sum_{i = 1}^{N}S_i   
			=\lim\limits_{\tau\rightarrow\infty}\frac{N}{\tau}\frac{1}{N}\sum_{i = 1}^{N}S_i = \Lambda E(S_i),$}	
	\end{equation}
where $S_i$ is the area of $i$-th triangle, $N$ refers to the number of reliable and secure packets, and $ \Lambda$ represents the average arrival rate of the packet over all time slots. In addition, $E(S_i)$ indicates the mean  of $S_i$, which can be given as	$E(S_i)=0.5E(A_i^2).$
	And $\Lambda$ can be expressed as $\Lambda=\frac{1}{E(A_i)+E(T_W)}$. The mean of $A_i$ can be calculated as 
	\begin{equation}\resizebox{0.98\hsize}{!}{$
				E(A_i)=\sum_{k = 1}^{\infty}(1-P_{ESP})^{k-1}P_{ESP}A_k=\frac{T}{(1-P_{B})(1-P_{d}(1-P_{E}))}.$}
	\end{equation}
The mean of $T_W$ can also be computed as 
	\begin{equation}\resizebox{0.65\hsize}{!}{$
				E(T_W)=\sum_{k = 1}^{\infty}(1-\lambda)^{k-1}\lambda T =\frac{T}{\lambda}.$}	
	\end{equation}
By combining the above results, we can obtain the average arrival rate of the packet as $\Lambda=\frac{1}{E(A_i)+E(T_W)}=\frac{P_{ESP}\lambda}{T(P_{ESP}+\lambda)}.$
	Likewise, $E(S_i)$ can be derived as 
	\begin{equation}\resizebox{0.75\hsize}{!}{$
			\begin{aligned}
				E(S_i)& =\frac{1}{2}E(A_i^2)=\frac{1}{2}\sum_{k = 1}^{\infty}(1-P_{ESP})^{k-1}P_{ESP}A_k^2  \\
				&=\frac{2T^2-(1-P_{B})(1-P_{d}(1-P_{E}))T^2}{2((1-P_{B})(1-P_{d}(1-P_{E})))^2}.
			\end{aligned}$}		\end{equation}
	By substituting the results into (7), the average SL will be 
	\begin{equation}\resizebox{0.65\hsize}{!}{$
		\begin{aligned}
			\overline{SL}&=\frac{\lambda}{\lambda+P_{ESP}}(\frac{T}{P_{ESP}}-\frac{T}{2}) \\ &=\frac{\lambda}{\lambda+(1-P_{B})(1-P_{d}(1-P_{E}))}\\ &\cdot(\frac{T}{(1-P_{B})(1-P_{d}(1-P_{E}))}-\frac{T}{2}).
		\end{aligned}$}	
	\end{equation}	
	 Based on (11), we have the following insights. Firstly, the SNR $\gamma$, block-length $L$, time slot $T$, and generating rate $\lambda$ influence the average SL jointly. Secondly, The average SL can be further written as $\overline{SL}=\frac{\lambda}{\lambda+P_{ESP}}(\frac{D}{R_{ES}}-\frac{T}{2})$, where the weighted proportion of the transmission time, i.e., $\frac{\lambda}{\lambda+P_{ESP}}$, represents the system workload for the transmitter. Thirdly, average SL can be simplified as the traditional average latency without security constraint, which is denoted as $\overline{L}=\frac{\lambda}{\lambda+(1-P_{B})}(\frac{T}{(1-P_{B})}-\frac{T}{2})$.
	\vspace{-5 pt}
	\section{Optimal Transmission Design Analysis}	
	In Section III, we derive the expression for the average SL. However, how to minimize the average SL by allocating the block-length and SNR is another critical issue while applied, which is analyzed in this section.
	\vspace{-10 pt}
	\subsection{Problem Reformulation}
	The optimization problem to minimize the average SL can be expressed as \vspace{-5pt}
	\begin{align}
		\underset{\gamma, L}{\text{minimize}}&~\quad\quad\overline{SL} \label{Problem}\\ 
		\text {s.t.}&~ D<L<L_{\text{max}},L\in\mathbb{N}^+, \tag{\ref{Problem}{a}}
	\end{align}
where (12a) represent the positive integer $L$ is limited by $D$ and the maximum value $L_{\text{max}}$.  Because of the discrete variable $L$ and continuous variable $\gamma$, and the nonlinearity induced by Q-functions and logarithmic terms, the optimization problem is a mixed-integer nonlinear programming problem. We noticed that only $P_{ESP}$ is related to $\gamma$ and $L$ in (11). Therefore, we calculate the first derivative of $\overline{SL}$ with respect to $P_{ESP}$ as \vspace{-3pt}
	\begin{equation}
		\begin{aligned}
			&~\frac{\alpha \overline{SL}}{\alpha P_{ESP}}
			=\frac{\alpha (\frac{\lambda}{\lambda+P_{ESP}}(\frac{T}{P_{ESP}}-\frac{T}{2}))}{\alpha P_{ESP}}      \\
			&~=\frac{-\lambda T((\lambda+P_{ESP})P_{ESP}+(2-P_{ESP})(2P_{ESP}+\lambda))}{2(\lambda+P_{ESP})^2{P_{ESP}}^2}  	.
		\end{aligned}
	\end{equation}
	Because of $\lambda$ $\in$ (0,1), the value of $\frac{\alpha\overline{SL}}{\alpha P_{ESP}}$ is negative, which means a bigger $P_{ESP}$ leads to a lower latency. Besides, we find that the optimal SNR and block-length for $P_{ESP}$ are also the optimal solutions for the average SL. We note $P_{ESP} \geq P^*_{ESP}$, where $P^*_{ESP}=(1-P_{B})P_{E}$. Here the equal sign holds when $P_{d}=1$.
	In addition, we find that $P_{ESP}$ is jointly influenced by block-length and SNR. Therefore, we divide the optimization problem into two subproblems, i.e., the optimal block-length optimization when SNR is determined, and the optimal SNR optimization when block-length is determined. Finally, the algorithm for solving problem (12) is presented.
\vspace{-10 pt}
	\subsection{Optimal Block-length and SNR Analysis}	
	For determined $\gamma$ and $D$, we can express the optimization problem as 
	\begin{equation}
		\begin{aligned}
			\underset{L}{\text{minimize}}&~~~~~~P_{ESP} \\
			\text {s.t.}& ~~~~~~ (12\text{a}).
		\end{aligned}	
	\end{equation}
	\begin{theorem} \label{theorem:1}
		For a small $\gamma$, the implicit solution of the optimal $L$ is shown in (15).
		\begin{figure*}
			\begin{equation}\resizebox{0.8\hsize}{!}{$
					\begin{aligned}	
						&~\exp(-\frac{L(\log(1+\gamma)-\frac{D}{L})^2}{2\gamma(\gamma+1)^{-2}(\gamma+2)})(1-\sqrt{L(\ln(1+\gamma)-\frac{\gamma}{\gamma+1})}(\frac{1}{2}+\frac{(-1)^i}{\sqrt{2\pi}}\sum_{i=0}^{n}\frac{(\sqrt{L}(1+\gamma)(\log(1+\gamma)-\frac{D}{L}))^{2i+1}}{i!2^i(2i+1)(\sqrt{\gamma(\gamma+2)})^{2i+1}})) \cdot \\
						&~\frac{2(1+\gamma)(\log(1+\gamma)L+D)}{L\sqrt{2\pi\gamma(\gamma+2)}}=(\frac{1}{2}+\frac{1}{\sqrt{2\pi}}\sum_{i=0}^{n}(-1)^i\frac{(\sqrt{L}(1+\gamma)(\log(1+\gamma)-\frac{D}{L}))^{2i+1}}{i!2^i(2i+1)(\sqrt{\gamma(\gamma+2)})^{2i+1}})^2\sqrt{\ln(1+\gamma)-\frac{\gamma}{\gamma+1}}
				\end{aligned}$}
			\end{equation}\vspace{-10pt}
		\end{figure*}
		For a large $\gamma$, the optimal $L$ can be derived as $\frac{D}{\log(1+\gamma)}$. The threshold $\gamma_{t}$ is the value  that satisfies the equation $\frac{\gamma_{t}}{(\gamma_{t}+1)\ln(\gamma_{t}+1)}=\frac{D\ln2-4}{D\ln2}$.
	\end{theorem}

	\begin{IEEEproof}
		See Appendix A.
	\end{IEEEproof}
	Theorem 1 provides the optimal transmission design with respect to block-length when SNR and $D$ are determined, which can be used in power-limited communications. We find that a smaller SNR or a bigger $D$ results in a longer optimal block-length. Besides, a higher SNR guarantees a lower packet error probability for both the receiver and eavesdropper, but also results in a higher detection probability. Similar to (14), for determined $L$ and $D$, the optimization problem can be formulated as 	
	\begin{equation}
		\begin{aligned}
			\underset{\gamma}{\text{minimize}}&~~~~~~P_{ESP} \\
			\text {s.t.}& ~~~~~~ (12\text{a}).
		\end{aligned} 	
	\end{equation}
	
	\begin{theorem} \label{theorem:2}
		For a small $L$, the implicit solution of the optimal $\gamma$ is shown in (17).
		\begin{figure*}
			\begin{equation}\resizebox{0.8\hsize}{!}{$
					\begin{aligned}
						&~\exp(\frac{L(\log(1+\gamma)-\frac{D}{L})^2}{-2\gamma(\gamma+1)^{-2}(\gamma+2)})(1-\sqrt{L(\ln(1+\gamma)-\frac{\gamma}{\gamma+1})}(\frac{1}{2}+\frac{(-1)^i}{\sqrt{2\pi}}\sum_{i = 0}^{n}\frac{(\sqrt{L}(1+\gamma)(\log(1+\gamma)-\frac{D}{L}))^{2i+1}}{i!2^i(2i+1)(\sqrt{\gamma(\gamma+2)})^{2i+1}})) \cdot \\
						&~\frac{\gamma(\gamma+2)-\ln2(\log(1+\gamma)-\frac{D}{L})}{\ln2(\gamma(\gamma+2))^\frac{3}{2}}=(\frac{1}{2}+\frac{(-1)^i}{\sqrt{2\pi}}\sum_{i = 0}^{n}\frac{(\sqrt{L}(1+\gamma)(\log(1+\gamma)-\frac{D}{L}))^{2i+1}}{i!2^i(2i+1)(\sqrt{\gamma(\gamma+2)})^{2i+1}})^2\frac{\gamma(\gamma+1)^{-\frac{3}{2}}}{4\sqrt{(\gamma+1)\ln(\gamma+1)-\gamma}}	
				\end{aligned}$}
			\end{equation} \vspace{-20pt}
		\end{figure*}
		For a large $L$, the optimal $\gamma$ can be derived as $2^{D/L}-1$. The threshold $L_t$ is the value that satisfies the equation $L_t(1-2^{-{D/L_t}})=D/\log e-4$.
	\end{theorem}
	
	\begin{IEEEproof}
		See Appendix B.
	\end{IEEEproof}
	Theorem 2 provides the optimal transmission design with respect to SNR when $L$ and $D$ are determined, which is applicable for the fixed block-length transmission to determine the optimal SNR based on the determined $L$ and $D$. 
	We note that the bigger $P_{ESP}$, the lower average SL, and that $P_{ESP} \geq P^*_{ESP}$, so the optimal $L$ and optimal SNR is from (15) and (17).
	Besides, we can find the peak of $P_{ESP}$ keeps increasing with the SNR and block-length when $P_{ESP} > \max(P^*_{ESP})$, i.e., $\frac{1}{4}$, although the curve of $P_{ESP}$ is non-convex in Fig. 4 \cite{zhuyao}. Therefore, the optimal $P_{ESP}$ can be achieved by iterating the optimal $L$ in (15) and optimal SNR in (17).
	\vspace{-15 pt}
	\section{Simulation Results} 
	\vspace{-2pt}
	In this section, the validity of the analysis is verified through numerical results. The impacts on the ESP and the averaged SL of SNR, block-length, and generating rate of the packet are also analyzed. In the simulations, the parameters are set as $D=64$ bits, noise power $\sigma_b^2=\sigma_e^2=-114$ dBm, $T=1/120$ ms.
	\begin{figure}[ht]\vspace{-7pt}
		\centering	
		\includegraphics[height=4.0cm,width=4.34cm]{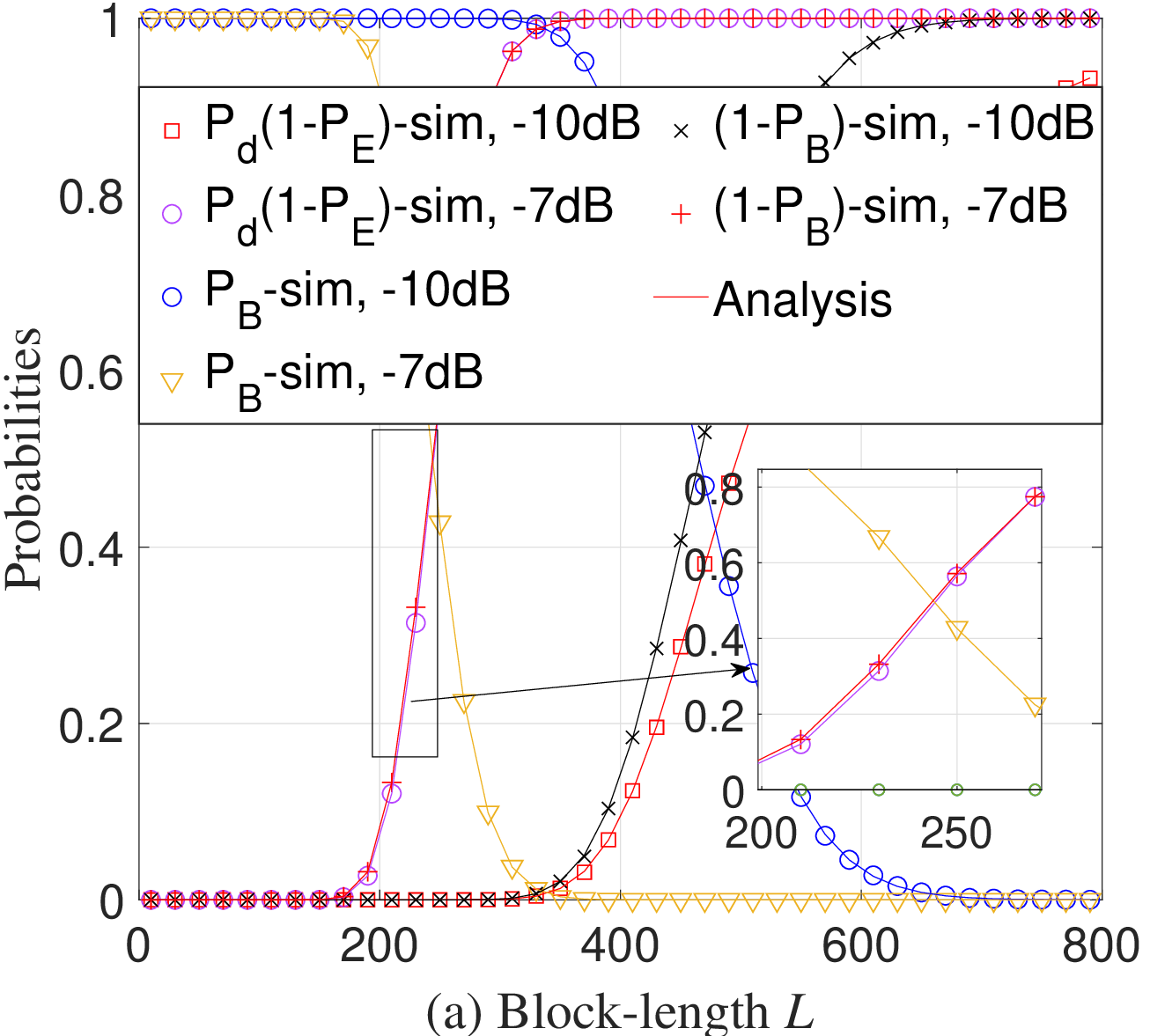}
		\includegraphics[height=4.0cm,width=4.34cm]{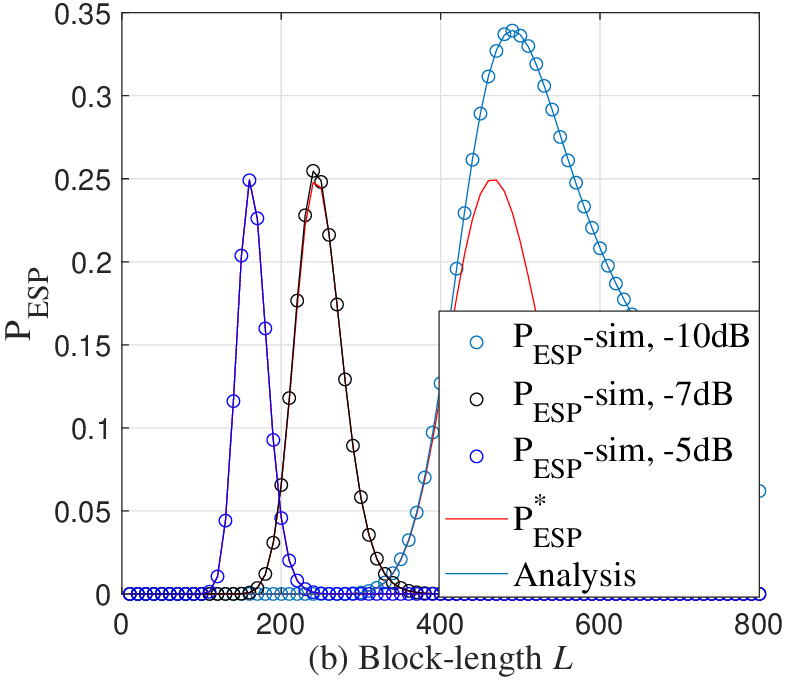}		
		\label{fig:side:b}	\vspace{-10 pt}
		\caption{(a) The probabilities of Bob's decoding and Eve's detecting and decoding $P_{d}(1-P_{E})$) versus the block-length. (b) $P_{ESP}$ versus the block-length.}\vspace{-10 pt}	
	\end{figure}
	In Fig. 2, we can see that when block-length is short, ESP is small. This is because when packets are shorter than the threshold, the reliability performance of the communication is inferior, which means that Bob's decoding is more likely to fail, although there is great security performance that decoding information is also difficult for Eve. This trend alleviates as the block-length increases. In contrast, ESP reduces as the block-length becomes longer than the threshold where Eve's performance equals Bob's reliability. This is because whenever Bob can easily decode the information, Eve can also decode easily, as shown in Fig.~2(a). The results indicate the trade-off between reliability and security. In addition, as the block-length grows, the detection probability at Eve $P_{d}$ approaches $1$, and thus the security constraint $P_{d}(1-P_{E})$ becomes $(1-P_{E})$, as shown in Fig. 2(a). Moreover, we can check from Fig. 2(b) that as the SNR increases, $P_{ESP}$ gradually approaches $P^*_{ESP}$, which is consistent with our analysis. It is particularly noteworthy that when the SNR is larger than the threshold, i.e., $-6.743$ dB, two curves almost overlap, which means that for the SNR that is larger than the threshold, we can take a simple form of optimal block-length $L=\frac{D}{\log(1+\gamma)}$. However, if SNR is smaller than the threshold, we have to take the optimal block-length $L$ in (15). We can verify that both $\gamma=-7$ dB and $\gamma=-10$ dB are lower than the threshold. So we calculated the optimal block-lengths to be $243$ bits and $489$ bits by employing the search algorithm in (15). When SNR is $-5$ dB, which is higher than the threshold, the optimal block-length is $161$ bits. The results from Theorem 1 are consistent with the simulations, which demonstrates the validness of our derivations therein.

	\begin{figure}[ht]\vspace{-5pt}
		\centering 
		\includegraphics[height=4.0cm,width=4.34cm]{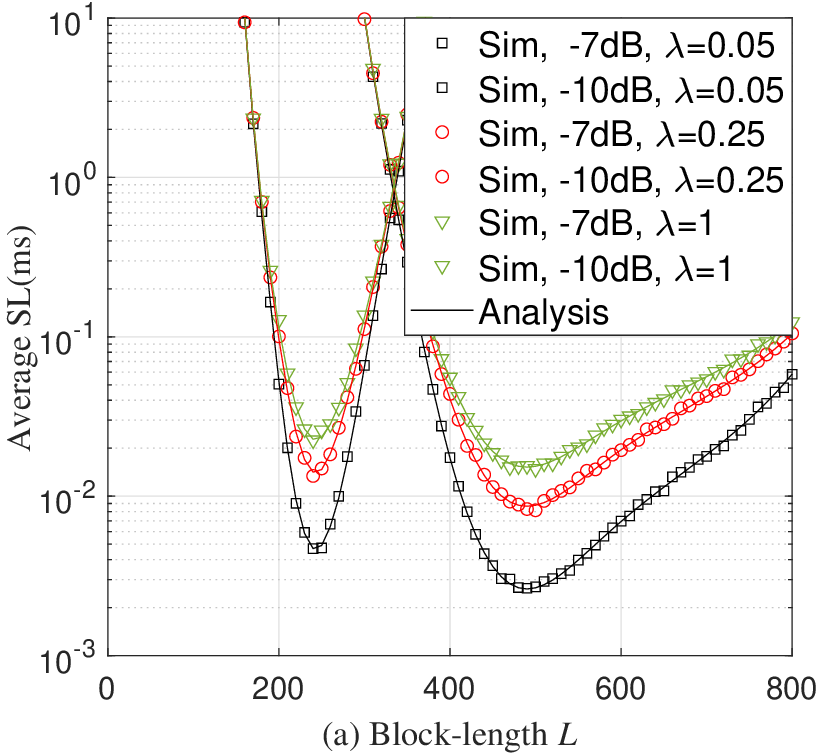}
		\includegraphics[height=4.0cm,width=4.34cm]{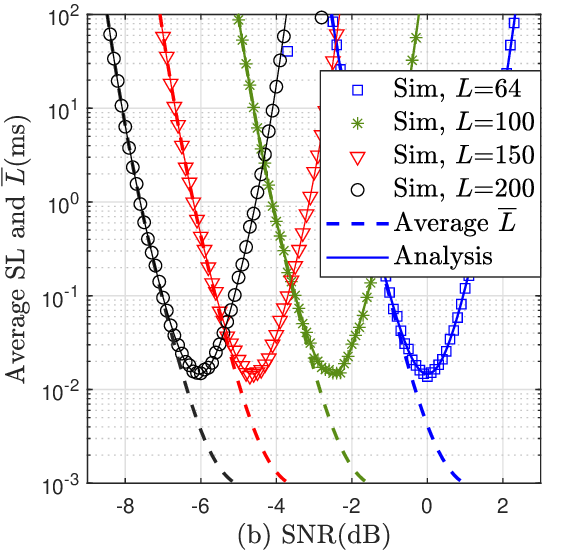}	
		\caption{ (a) Average SL versus the block-length with different $\lambda$ and SNR. (b) Average SL and $\overline{L}$ versus SNR. }\vspace{-10 pt}
		\label{fig}	
	\end{figure}
	
	In Fig. 3(a), we notice that there exists a minimum average SL. This is because when the packet is shorter than the threshold, a larger $L$ results in a larger ESP, leading to reduced retransmissions and a decreased latency. Average SL afterwards grows with block-length. This is because a longer block-length makes information easier to be detected and decoded by Eve. Additionally, we find that a higher $\lambda$ results in a higher average SL because the higher $\lambda$, the more frequent transmissions.  
	
	Fig. 3(b) shows the theoretical analysis of average SL and $\overline{L}$ versus SNR with different $L$ and $D$. We find that there is a minimum average SL in terms of SNR. In addition, the optimal SNR changes along with block-length. Similar to the validation of Theorem 1, when packet lengths are $64$ bits, $100$ bits, $150$ bits and $200$ bits, the optimal SNR are $0$ dB, $-2.531$ dB, $-4.633$ dB and $-6.050$ dB respectively according to Theorem 2, which are consistent with the simulation results. In addition, compared to the average $\overline{L}$ that only involves reliability, the average SL increases with SNR increasing. This is because as SNR increases, although reliability is guaranteed, security risks also increase, which indicates the trade-off between reliability and security.

    Fig. 4 shows the $P_{ESP}$ versus SNR and the block-length. The mark represents the result from the iterating algorithm, which is consistent with the biggest $P_{ESP}$. Because the value of $\frac{\alpha\overline{SL}}{\alpha P_{ESP}}$ is negative, the optimal solution of problem (12) is solved by iterating the optimal solutions in Theorem 1 and Theorem 2. In addition, we find that the optimal block-length is the longest, and that the optimal SNR can be obtained by applying the search algorithm in (17).

		\begin{figure}[ht]
		\centering\vspace{-5pt}
		\includegraphics[height=3.0cm,width=5.0cm]{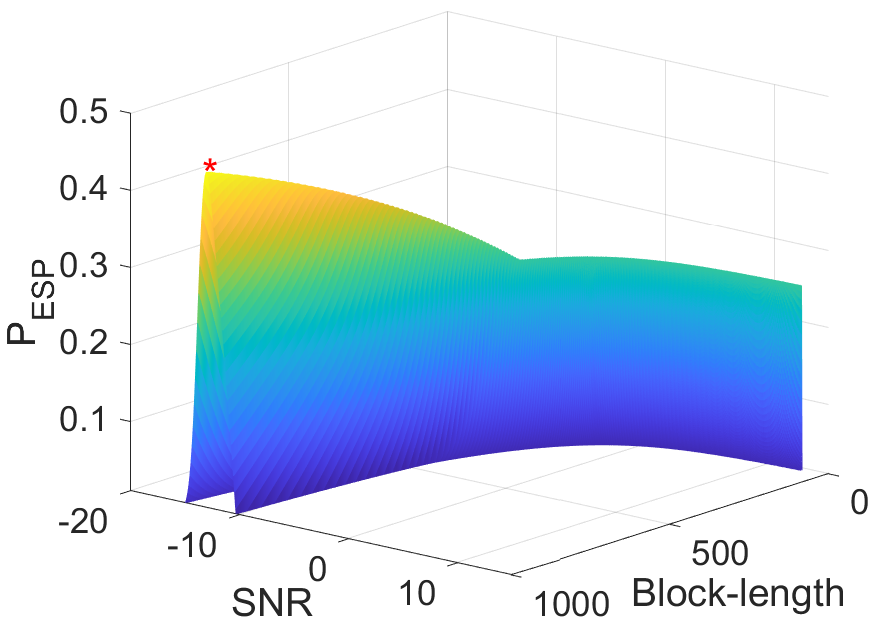}
		\label{fig:side:g}	
		\caption{$P_{ESP}$ versus SNR and block-length.}	\vspace{-10 pt}
	\end{figure}
	\vspace{-5pt}

	\section{Conclusion} 
	
	In this article, we have investigated the design to minimize the security-latency performance by allocating the SNR and block length in which we have proposed ESP and average SL to reveal the impact of SNR and the block-length on latency in short packet-based low-altitude communications. To solve the optimization problem, we have analyzed different designs in which the optimal analytical solutions of the block-length and SNR are given. The simulation results have verified the accuracy of the analysis and revealed the trade-off between reliability and security and the impacts of block-length, SNR, and packet generation rate on average SL, of which the block-length and SNR are the main factors. We have found the logest block-length enables the minimized average SL and that the security-latency performance can be enhanced by allocating less SNR.
 
\vspace{-9pt}
	\section*{Appendix A \\	 Proof of Theorem~\ref{theorem:1} }	
	The first derivative of $P_{ESP}$ over $L$ can be obtained as
	\begin{equation}
			\begin{aligned}	
				&~\frac{\partial P_{ESP}}{\partial L}=\frac{\partial ((1-P_{e})(1-P_{d}(1-P_{e})))}{\partial L}\\
				&~=\frac{\partial P_{e}}{\partial L}(2P_{d}(1-P_{e})-1)-(1-P_{e})^2\frac{\partial P_{d}}{\partial L}.
		\end{aligned}	
	\end{equation}
	According to (3), $\frac{\partial P_{e}}{\partial L}$ can be expressed as
$\frac{\partial P_{e}}{\partial t}\frac{\partial t}{\partial L}
				=-\frac{(1+\gamma)(\log(1+\gamma)L+D)}{2\sqrt{2\pi\gamma(\gamma+2)L^3}}\exp(\frac{L(\log(1+\gamma)-D/L)^2}{-2\gamma(\gamma+1)^{-2}(\gamma+2)}),$
	where $t=\frac{C(\gamma)-R}{\sqrt{V(\gamma)/L}}$. Then $\frac{\partial P_{d}}{\partial L}$ can be derived as $
				\frac{\partial P_{d}}{\partial L}=\sqrt{(\ln(1+\gamma)-\frac{\gamma}{\gamma+1})/(16L)}.$
	We apply Taylor series to the Q function in $P_{e}$, which gives $
				1-P_{e}=\frac{1}{\sqrt{2\pi}}\int_{-\infty}^{t}\exp({-x^2}/{2}) dx=\frac{1}{2}+
				\frac{1}{\sqrt{2\pi}}\sum_{i = 0}^{\infty}(-1)^i\frac{[\sqrt{L}(1+\gamma)(\ln(1+\gamma)-D/L)]^{2i+1}}{i!2^i(2i+1)[\sqrt{\gamma(\gamma+2)}]^{2i+1}}.$
	By substituting the above results into (18), letting $\frac{\partial (P_{ESP})}{\partial L}=0$ and applying the search algorithm to the implicit solution in (15), the optimal $L$ is obtained.
	For a higher SNR, $P_{d}$ approaches $1$ as block-length increase, i.e., Eve always detects the transmission. Therefore, $P_{ESP}$ approaches $P^*_{ESP}$. So the first derivative of $P_{ESP}$ with respect to $L$ will be $\frac{\partial P_{ESP}}{\partial L}=\frac{\partial(1-P_{e})P_{e}}{\partial L}=\frac{\partial P_{e}}{\partial L}(1-2P_{e}).$
	Because $\frac{\partial P_{e}}{\partial L}$ is negative, $P_{e}$ should equal $0.5$ to make $\frac{\partial P_{ESP}}{\partial L}=0$.	Then we have $P_{e} \approx Q(\frac{C(\gamma)-R}{\sqrt{V(\gamma)/L}})=Q(0).	$
	Therefore, for a higher SNR, the optimal $L$ is obtained as $\frac{D}{\log(1+\gamma)}$. And the threshold $\gamma_{t}$ to distinguish two cases is whether the optimal $L$ makes $P_{d}$ equal $1$, which means $\sqrt{\frac{D}{4\log(1+\gamma_{t})}(\ln(\gamma_{t}+1)-\frac{\gamma_{t}}{\gamma_{t}+1})}=1$. Finally, $\gamma_{t}$ satisfies $\frac{\gamma_{t}}{(\gamma_{t}+1)\ln(\gamma_{t}+1)}=\frac{D\ln2-4}{D\ln2}$.

	\section*{Appendix B \\	 Proof of Theorem~\ref{theorem:2} }	
	
	Similarly, the first derivative of $P_{ESP}$ with respect to $\gamma$ can be derived as
	\begin{equation}
		\begin{aligned}
			&~\frac{\partial P_{ESP}}{\partial \gamma}=\frac{\partial (1-P_{e})(1-P_{d}(1-P_{e}))}{\partial \gamma}\\
			&~=\frac{\partial P_{e}}{\partial \gamma}(2P_{d}(1-P_{e})-1)-(1-P_{e})^2\frac{\partial(P_{d})}{\partial \gamma}.
		\end{aligned}	
	\end{equation}
	According to (3), $\frac{\partial P_{e}}{\partial \gamma}$ can be obtained as $\frac{\partial P_{e}}{\partial t}\frac{\partial t}{\partial \gamma}=\frac{\ln2(\log(1+\gamma)+D/L)-\gamma(\gamma+2)}{L^{-1/2}\ln2[\gamma(\gamma+2)]^{3/2}}\exp(\frac{L(\ln(1+\gamma)-D/L)^2}{-2\gamma(\gamma+1)^{-2}(\gamma+2)}).$
	In addition, $\frac{\partial P_{d}}{\partial \gamma}$ can be calculated as $\frac{\partial P_{d}}{\partial \gamma}=\frac{\sqrt{L}\gamma(\gamma+1)^{-3/2}}{4\sqrt{(\gamma+1)\ln(\gamma+1)-\gamma}}.$
	By substituting the above results into (19), letting $\frac{\partial (P_{ESP})}{\partial \gamma}=0$ and applying the search algorithm to the implicit solution in (17), the optimal $\gamma$ is obtained. For a longer block-length, $P_{d}$ approaches $1$ as SNR increases. In this case, $P_{ESP}=P^*_{ESP}=(1-P_{e})P_{e} \leqslant 0.25$, with equality when $P_{e}=0.5$. Similarly, the optimal $\gamma$ is derived as $\gamma=2^{D/L}-1$. And the threshold for distinguishing two cases $L_t$ satisfies $L_t(1-2^{-D/L_t})=D/\log e-4$.

	\bibliographystyle{IEEEtran}
	\bibliography{ices}
	
\end{document}